\documentclass[iop,apj,numberedappendix,appendixfloats]{emulateapj}

\usepackage{amsmath}
\usepackage{multirow}
\usepackage{graphicx,float}
\usepackage{gensymb}
\usepackage[flushleft]{threeparttable}

\bibpunct[, ]{(}{)}{;}{a}{}{,}

\newcommand{\Omegam}{\ensuremath{\Omega_{\mathrm{m}}}}

\newcommand{\hmvol}{h^{-3}{\mathrm{Mpc}}^{3}}

\newcommand{\hmpc}{h^{-1}\mathrm{Mpc}}

\newcommand{\lcdm}{\ensuremath{\Lambda\mathrm{CDM}}}

\begin{document}

\title{A Cosmic Void Catalog of SDSS DR12 BOSS Galaxies}

\author{
  Qingqing~Mao\altaffilmark{1},
  Andreas~A.~Berlind\altaffilmark{1,2},
  Robert~J.~Scherrer\altaffilmark{1},
  Mark~C.~Neyrinck\altaffilmark{3},
  Rom\'an~Scoccimarro\altaffilmark{4},
  Jeremy~L.~Tinker\altaffilmark{4},
  Cameron~K.~McBride\altaffilmark{5},
  Donald~P.~Schneider\altaffilmark{6},
  Kaike~Pan\altaffilmark{7},
  Dmitry~Bizyaev\altaffilmark{7,8},
  Elena~Malanushenko\altaffilmark{7},
  Viktor~Malanushenko\altaffilmark{7}
}
\altaffiltext{1}{Department of Physics and Astronomy, Vanderbilt University, Nashville, TN 37240, USA}
\altaffiltext{2}{a.berlind@vanderbilt.edu}
\altaffiltext{3}{Department of Physics and Astronomy, The Johns Hopkins University, Baltimore, MD 21218, USA}
\altaffiltext{4}{Center for Cosmology and Particle Physics \& Department of Physics, New York University, New York, NY 10003, USA}
\altaffiltext{5}{Harvard-Smithsonian Center for Astrophysics, Cambridge, MA 02138, USA}
\altaffiltext{6}{Institute for Gravitation and the Cosmos, Department of Astronomy and Astrophysics, The Pennsylvania State University, University Park, PA 16802, USA}
\altaffiltext{7}{Apache Point Observatory and New Mexico State University, P.O. Box 59, Sunspot, NM, 88349-0059, USA}
\altaffiltext{8}{Sternberg Astronomical Institute, Moscow State University, Moscow, Russia}

\begin{abstract}
We present a cosmic void catalog using the large-scale structure galaxy catalog from the Baryon Oscillation Spectroscopic Survey (BOSS). This galaxy catalog is part of the Sloan Digital Sky Survey (SDSS) Data Release 12 and is the final catalog of SDSS-III. We take into account the survey boundaries, masks, and angular and radial selection functions, and apply the ZOBOV void finding algorithm to the galaxy catalog. We identify a total of 10,643 voids. After making quality cuts to ensure that the voids represent real underdense regions, we obtain 1,228 voids with effective radii spanning the range 20-100$\hmpc$ and with central densities that are, on average, 30\% of the mean sample density. We release versions of the catalogs both with and without quality cuts. We discuss the basic statistics of voids, such as their size and redshift distributions, and measure the radial density profile of the voids via a stacking technique. In addition, we construct mock void catalogs from 1000 mock galaxy catalogs, and find that the properties of BOSS voids are in good agreement with those in the mock catalogs. We compare the stellar mass distribution of galaxies living inside and outside of the voids, and find no significant difference. These BOSS and mock void catalogs are useful for a number of cosmological and galaxy environment studies.

\end{abstract}


\section{Introduction}
Cosmic voids are large underdense regions present in the hierarchical structure of the Universe. Surrounded by filaments, walls and clusters, voids are an essential component of the cosmic web. They were first discovered in early galaxy redshift surveys \citep{Gregory:1978, Kirshner:1981, Lapparent:1986} over thirty years ago. More recent redshift surveys such as the 2dF Galaxy Redshift Survey (2dFGRS; \citealt{Colless:2001}) and the Sloan Digital Sky Survey (SDSS; \citealt{York:2000}), have greatly expanded our view of the large-scale structure, and provide much larger data sets to study void properties systematically and in detail. 

Cosmic voids have been recognized as interesting cosmological laboratories for investigating galaxy evolution, structure formation and cosmology. The low-density environment of voids provides an ideal place to examine the influence of environment on the formation and evolution of galaxies \citep{Peebles:2001, Gottlober:2003, Rojas:2004, Rojas:2005, Hoyle:2005, Hoyle:2012}. Voids also contain information on the structure formation history and cosmological scenario. The size and shape distribution of voids, their intrinsic structure, and their counts can provide insights into the growth of structure \citep{Jennings:2013} and dark energy \citep{Lee:2009, Biswas:2010, Bos:2012, Pisani:2015}. Moreover, the Alcock-Paczy\'nski test \citep{Alcock:1979} can be applied to ``stacked" voids to probe the expansion history of the universe \citep{Ryden:1995, Lavaux:2012, Sutter:2012b}. Voids can also be correlated with the cosmic microwave background \citep{Bennett:2013} to study the integrated Sachs-Wolfe effect \citep{Thompson:1987, Granett:2008, PlanckCollaboration:2014}. Since voids are nearly empty, the dynamics in their interior are dominated by dark energy \citep{Goldberg:2004}, making them potentially important probes for studying the nature of dark energy and testing exotic physics such as modified gravity or a fifth force \citep{Li:2012, Spolyar:2013, Clampitt:2013, Zivick:2014, Hamaus:2015}. 

To unleash the power of these cosmological applications, it is important to first find voids robustly from simulations, mock galaxy catalogs and galaxy surveys. Although voids occupy most of the volume in the Universe, they are not straightforward to define and identify, especially in surveys where the density field is traced by a set of sparsely sampled galaxies and the survey geometry is complicated. There exist a number of quite different void-finding algorithms \citep{Colberg:2008}. While each algorithm has different advantages and disadvantages, \citet{Colberg:2008} found that their basic results agree with each other when applied to the dark matter distributions of N-body simulations. One popular algorithm among these is ZOBOV \citep{Neyrinck:2008}, which is based on Voronoi tessellations and the watershed method \citep{Platen:2007}. One of the advantages of ZOBOV is that it does not assume anything about void shape, thus allowing us a full exploration of the natural shape of voids and their hierarchical structure. ZOBOV in general is parameter free, but additional restrictions can be introduced as needed. 

In this paper we present a catalog of voids by applying the ZOBOV algorithm to SDSS data. There have been previous void catalogs produced from the SDSS data. \citet{Pan:2012} identified voids in the SDSS Data Release 7 (DR7; \citealt{Abazajian:2009}) main galaxy sample \citep{Strauss:2002} using a nearest neighbor algorithm. Recently, \citet{Sutter:2012a} successfully applied ZOBOV to the SDSS DR7 main galaxy sample and the luminous red galaxy sample \citep{Eisenstein:2001}, and \citet{Sutter:2014} applied ZOBOV to the SDSS Data Release 9 (DR9; \citealt{Ahn:2012}) CMASS sample. We apply ZOBOV to the most recent SDSS Data Release 12 (DR12; \citealt{Alam:2015}) CMASS and LOWZ galaxy samples, which comprise the largest spectroscopic galaxy redshift samples available to date. We take into account the survey geometry and completeness. The void catalogs will be useful for many void-based studies in cosmology and galaxy formation and evolution. 

In \S\ref{s:data}, we describe the galaxy and mock catalogs used in this study. In \S\ref{s:method}, we present the void finding methodology in detail. We describe the resulting void catalogs in \S\ref{s:catalog} and show statistics of the identified voids in \S\ref{s:statistics}. Conclusions and discussion follow in \S\ref{s:conclusion}.

\section{LSS catalog and QPM mocks}
\label{s:data}
The galaxy sample used in this study is from the Baryon Oscillation Spectroscopic Survey (BOSS; \citealt{Dawson:2013}), which is part of the third generation of the Sloan Digital Sky Survey (SDSS-III; \citealt{Eisenstein:2011}). BOSS made use of the dedicated SDSS telescope \citep{Gunn:2006}, multi-object spectrograph \citep{Smee:2013}, and software pipeline \citep{Bolton:2012}, to obtain the spectra of over 1.37 million galaxies over two large contiguous regions of sky in the Northern and Southern Galactic Caps, covering over 10,000 $\mathrm{deg}^2$ in total. DR12 is the final data release of SDSS-III and contains all six years of BOSS data. 

We use the large-scale structure (LSS) galaxy catalogs for DR12 produced by the BOSS collaboration. BOSS galaxies were uniformly targeted in two samples: a relatively low-redshift sample with $z < 0.45$ (LOWZ) and a sample with $0.4 < z < 0.7$ that was designed to be approximately volume-limited in stellar mass (CMASS). A full description of the targeting criteria can be found in \citet{Dawson:2013}. We place the redshift cuts $0.2 < z < 0.43$ on the LOWZ sample and $0.43 < z < 0.7$ on the CMASS sample to ensure clear geometric boundaries and no overlap between samples. Our study has four large areas of data, CMASS North and South, and LOWZ North and South. 

Due to hardware constraints and pipeline failures, not all targeted galaxies result in a good redshift measurement. Each galaxy is weighted to correct for the effects of redshift failures and fiber collisions (no two targets in a spectroscopic observation can be separated by less than 62$\arcsec$ on the sky). In addition, there are weights to account for the systematic relationships between the number density of observed galaxies and stellar density and seeing. These weights are all included in the LSS catalogs and their detailed description can be found in Reid et al. (in preparation). 

The LSS catalogs use the MANGLE software \citep{Swanson:2008} to account for the survey geometry and angular completeness. For each distinct region, we up-weight all the galaxies in the region according the its completeness to correct for the angular selection function. The LOWZ and CMASS samples are not strictly volume-limited and their number densities depend on redshift. This redshift dependence of density does not strongly impact void properties because most voids do not span a wide enough redshift range to be sensitive to changes in the underlying density. However, anytime we need to compare a local density measurement to the mean density of the sample, we always compare it to the observed radial density distribution $n(z)$, measured at the corresponding redshift.

To test our void finding algorithm, we also use a set of 1,000 mock galaxy catalogs generated using the ``quick particle mesh'' (QPM) methodology described by \citet{White:2014}. These QPM mocks were based on a set of low-resolution particle mesh simulations that accurately reproduce the large-scale dark matter density field on few Mpc scales. Each simulation contained $1280^3$ particles in a box of side length 2,560 $\hmpc$. The chosen cosmology has $\Omegam = 0.29$, $h = 0.7$, $n_s = 0.97$ and $\sigma_8 = 0.8$. Mock halos were selected based on the local density of each particle, and populated using the halo occupation distribution \citep[HOD; e.g.,][]{Berlind:2002} method to create galaxy mocks. The HOD was chosen such that the clustering amplitude of mock galaxies matches the observed measurements. The survey masks were then applied so that the mock catalogs have the same survey geometry as the BOSS data. Finally, the mock catalogs were randomly down-sampled to have the same angular sky completeness and the same radial mean $n(z)$ as the data. We have mock catalogs for the CMASS North and South samples, but not the LOWZ samples.

\begin{figure*}
	\includegraphics[scale=1]{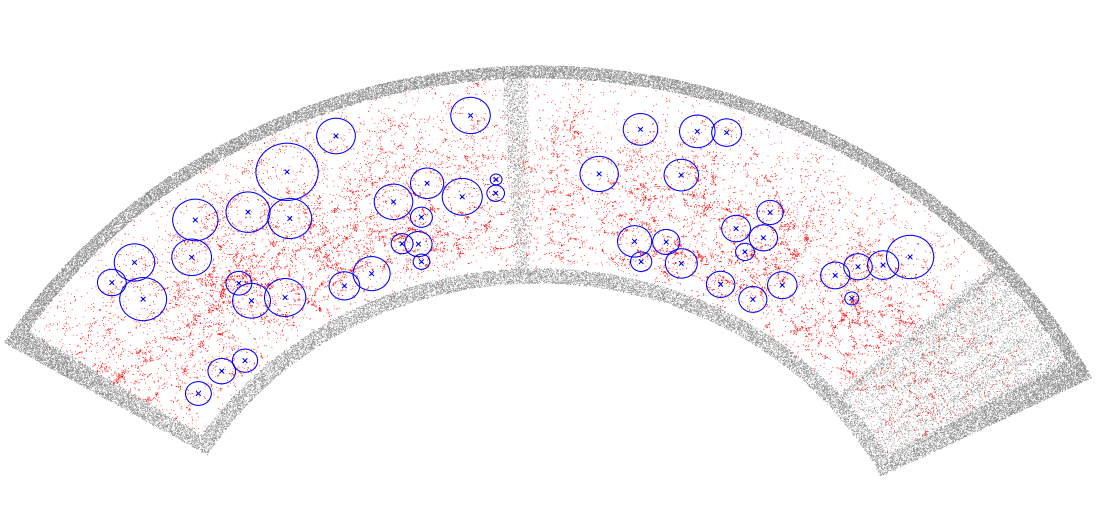}
	\caption{A thin slice of CMASS North galaxies (red) and random buffer particles (gray). The slice is centered on the celestial equator and is $2\degree$ thick in declination. Blue crosses show the central positions of the identified voids whose weighted centers are also located in the slice; the sizes of the blue circles indicate the  effective radii of the voids.}
	\label{fig:geometry_slice}
\end{figure*}

\section{Void finding algorithm}
\label{s:method}
We use the ZOBOV algorithm to find voids in the BOSS LSS catalog and QPM mock catalogs. The first step of ZOBOV is to perform a Voronoi tessellation on a given set of particles. The tessellation assigns each particle a Voronoi cell defined as the set of all points in space that are closer to that particle than to any other. The volume of the Voronoi cell provides a density estimate for each particle. The tessellation also provides a natural adjacency measurement for each particle. ZOBOV then applies the watershed transform algorithm to group neighboring Voronoi cells into zones and eventually subvoids and voids. Each void is like a basin composed of a set of attached Voronoi cells, surrounding a local density minimum. ZOBOV also measures the statistical significance of a void by comparing its density contrast, which is the ratio the density measured at the void ridge to the minimum density, to the distribution of density contrasts that can arise from Poisson fluctuations. For a more detailed discussion of this analysis package, see \citet{Neyrinck:2008}. 

To run the Voronoi tessellation, we first convert galaxy redshifts to line-of-sight distances assuming a flat $\lcdm$ universe with $\Omegam = 0.3$. We then prepare the LSS catalog to take into account the survey geometry. The survey masks are used to generate a high number density of randomly distributed buffer particles (at least ten times higher density than the BOSS galaxies) and place them just outside and all around the survey boundaries. The purpose of these buffer particles is to ensure the tessellation process works even for galaxies close to the survey boundaries. However, the buffer particles and the galaxies adjacent to buffer particles are not included in the watershed transform step. Figure~\ref{fig:geometry_slice} displays a thin slice of the galaxies from the CMASS North sample together with the buffer particles that surround the survey geometry and fill the holes. 

All the weights are applied immediately after the tessellation step by directly modifying the corresponding number density of each galaxy as $n_i = w_i / V_i$, where $w_i$ is the total weight of the galaxy and $V_i$ is the volume of the Voronoi cell. However, all the adjacency information is retained untouched. This is an easy way to include the systematic weights and apply the angular selection function. The watershed method can then be run smoothly with no additional modification. 

In general, ZOBOV can be parameter free, but some restrictions produce catalogs better-suited to typical void analysis. For example, ZOBOV zones and voids are grouped around all local minima, including those sitting in high density environments, in which case an identified void may actually have a high mean density. Since we are interested in low density regions, we set some density criteria during the void finding process. First, there is a density threshold parameter that limits ZOBOV to only group zones with mean density less than a certain level during the watershed transform step. This value is set to 0.5, which means that only zones with mean density lower than half the mean density of the whole sample can be joined to voids. In sparsely sampled catalogs, most physical voids only contain one zone, in which case this density threshold parameter has no effect. We also exclude voids where the minimum Voronoi density is higher than the mean density of the sample. Finally, only voids with significance larger than $2\sigma$ are included, which is calculated based on the depth of a void (see the next section for the detailed description). These last two cuts are meant to ensure that the resulting voids represent real physical underdense regions. However, some science applications might benefit from making less stringent cuts that result in a larger void sample. For this reason we also provide versions of void catalogs without these two cuts.

\section{Void catalogs}
\label{s:catalog}

\begin{table*}
  \centering
  \begin{threeparttable}
  	\centering 
  	\caption{Void catalog from the BOSS CMASS North sample.}
		\label{table:catalog}
		\begin{tabular}{rcccrccccccc}
  		\hline \hline
    	ID & RA (J2000) & {\centering}DEC (J2000)& $z$ & $N_\mathrm{gal}$ & $V$ & $R_\mathrm{eff}$ & $n_\mathrm{min}$ & $\delta_\mathrm{min}$ & $r$ & $P$ & $D_\mathrm{boundary}$ \\
		 & (deg) & (deg) &  &  & ($\hmvol$) & ($\hmpc$) & ($h^{3}{\mathrm{Mpc}}^{-3}$)  & & & & ($\hmpc$) \\
  (1) & (2) & (3) & (4) & (5) & (6) & (7) & (8) & (9) & (10) & (11) & (12) \\
 		\hline \hline
		60    &  114.782 & +37.641 & 0.648  & 35   &   1.411e+05  &     32.298 & 2.486e-05   &    $-0.717$ & 3.922 &  3.220e-14  &     52.504 \\
		10020 &  184.261 & +1.326  & 0.500  & 25   &   1.704e+04  &     15.964 & 1.364e-04   &    $-0.652$ & 3.441 &  2.200e-10  &     28.489 \\
		11496 &  124.855 & +3.090  & 0.648  & 117  &   6.052e+05  &     52.473 & 1.872e-05   &    $-0.778$ & 3.372 &  6.630e-10  &     54.891 \\
		15935 &  230.976 & +13.239 & 0.459  & 83   &   2.425e+05  &     38.683 & 3.120e-05   &    $-0.876$ & 3.328 &  1.330e-09  &     57.265 \\
		4407  &  237.406 & +16.985 & 0.463  & 372  &   1.071e+06  &     63.467 & 2.934e-05   &    $-0.884$ & 3.001 &  1.330e-07  &     73.644 \\
    	\hline \hline
		\end{tabular}
	\begin{tablenotes}
		\small
		\item Notes---The columns are described in the text. Table 1 is available in its entirety in the electronic version of this paper, as well as at http://lss.phy.vanderbilt.edu/voids.
	\end{tablenotes}
  \end{threeparttable}
\end{table*}

We apply the ZOBOV algorithm to four separate regions of BOSS galaxies: CMASS North, CMASS South, LOWZ North, and LOWZ South. Before making the two quality cuts described in the previous section, in these regions we find a total of 5734, 2010, 1983, and 916 voids, respectively. After making the cuts, we find 584, 190, 319, and 135 voids, respectively. We parse the ZOBOV outputs and calculate the essential properties for all the voids. For each void, we find the weighted center of the void, which is the average position of the void galaxies weighted by the inverse of their Voronoi density, 
\begin{equation}
	\mathbf{X} = \frac{\sum_i \mathbf{x}_i / n_i}{\sum_i 1 / n_i},
	\label{eq:voidcenter}
\end{equation}
where $\mathbf{x}_i$ are the positions of the galaxies in the void and $n_i$ are their corresponding Voronoi densities. The Voronoi density of each galaxy is defined as $n_i = w_i / V_i$, where $w_i$ is the weight of the galaxy and $V_i$ is the Voronoi volume from the tessellation. The effective radius of a void is defined as 
\begin{equation}
	R_\mathrm{eff} \equiv \left(\frac{3}{4\pi}V\right)^{1/3}, 
	\label{eq:Reff}
\end{equation}
where $V$ is the total Voronoi volume of the void, which is equal to the sum of Voronoi volumes of all the member galaxies in the void. We also provide the density minimum of the void, as well as its density contrast compared to the mean density at that redshift. ZOBOV calculates the ratio between the minimum density particle on a ridge to the minimum density particle of the whole void. This ratio, $r$, is used to determine the statistical significance of the void by comparing it to those arising from Poisson fluctuations. Both these measurements are given in our catalogs. Finally, we calculate the distance from each void's weighted center to its nearest survey boundary by finding the nearest buffer particle to the void center. 

In Table~\ref{table:catalog}, we present a few of the most significant voids in the CMASS North sample (the version with quality cuts). We list the void ID (col. [1]); the (J2000.0) right ascension and declination of the void weighted center (cols. [2] and [3]); the redshift of the weighted center (col. [4]); the number of galaxies in the void, $N_\mathrm{gal}$ (col. [5]); the total Voronoi volume of the void, $V$ (col. [6]); the effective radius, $R_\mathrm{eff}$ (col. [7]); the number density of the minimum density Voronoi cell in the void, $n_\mathrm{min}$ (col. [8]); the density contrast of the minimum density cell comparing to the mean density at that redshift, $\delta_\mathrm{min}$ (col. [9]); the ratio $r$ between the minimum density particle on a ridge to the minimum density particle of the void (col. [10]); the probability that the void arises from Poisson fluctuations (col. [11]); the distance from the weighted center to the nearest survey boundary (col. [12]). The voids are ranked in decreasing order of the probability. The complete void catalogs for all four galaxy samples are published in the electronic version of this article. These catalogs, the accompanying void galaxy membership lists, the complete lists of Voronoi volumes for all galaxies, and the void catalogs from the 1,000 mock catalogs are also available for download on an external site \footnote{http://lss.phy.vanderbilt.edu/voids}. We provide both the full uncut versions of the void and member catalogs, and the versions that include cuts to ensure that the voids are significant underdense regions. For the rest of this paper, we only show results using the version that includes cuts.

To visualize the voids, their positions are displayed in the slice in Figure~\ref{fig:geometry_slice}, with their effective radii indicated by the circles. Although the circles indicating the effective radii appear to overlap in some cases, the voids found by ZOBOV do not actually overlap with each other. A ZOBOV void either stands alone, or is fully embedded in a larger void as a subvoid. All the voids that satisfy our criteria in this catalog are stand-alone voids. Some of the voids identified in Figure~\ref{fig:geometry_slice} appear to contain high density regions. This effect is partly due to projection and partly due to the fact that ZOBOV voids are not actually spherical and so not all the region inside the effective radius is necessarily part of the void. We show that the voids in our catalogs actually represent underdense regions when we investigate their stacked density profiles in the next section.

\section{Void statistics and properties}
\label{s:statistics}

\begin{figure}
	\includegraphics[scale=0.4]{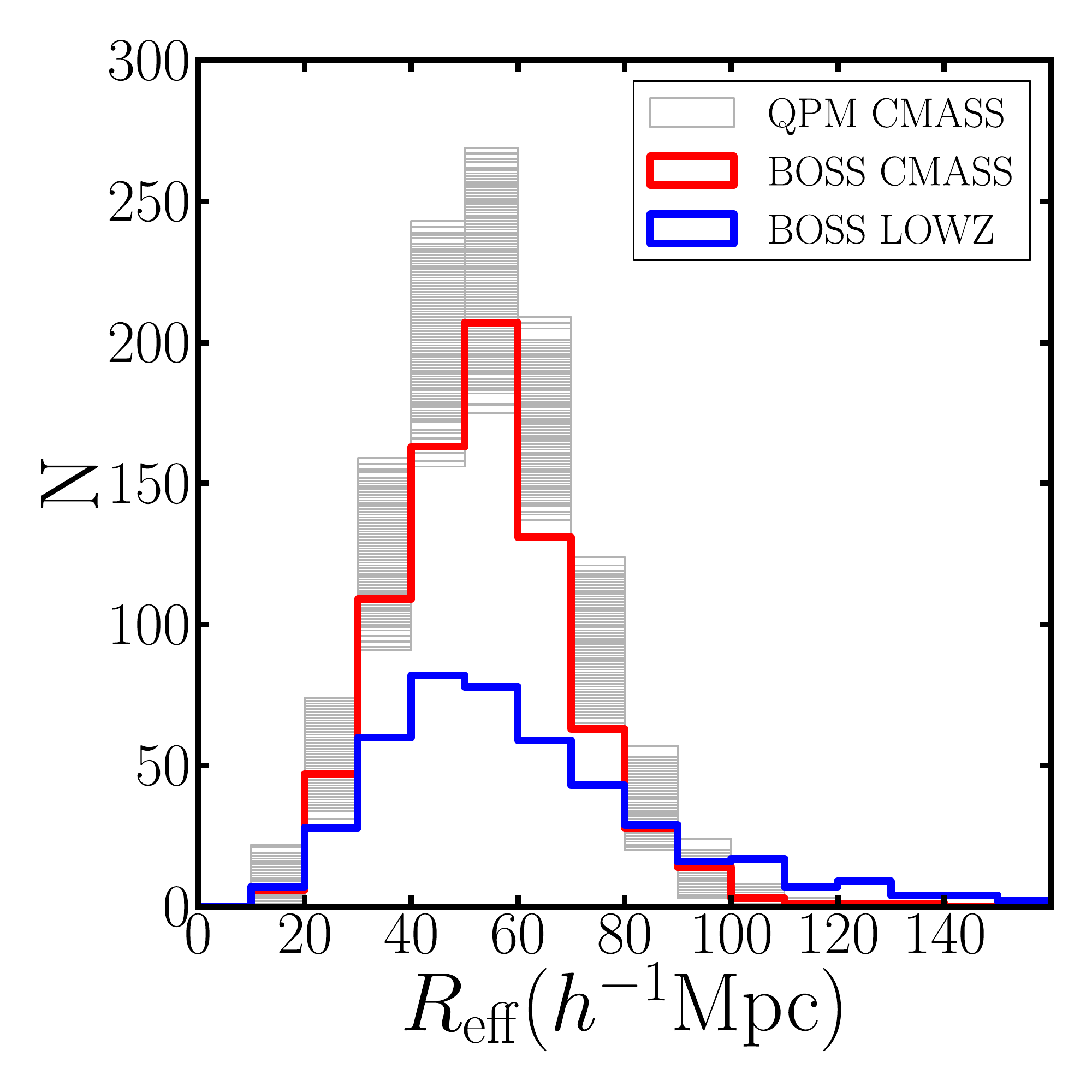}
	\caption{Distribution of void sizes. Each gray line represents the void effective radius distribution for one of the 1000 QPM CMASS (North + South) mock catalogs. Results for the CMASS (North + South) and LOWZ (North + South) samples are shown by the red and blue lines, respectively. Most voids have effective radii between $30$ and $80\hmpc$. The mock catalogs contain, on average, 10-20\% more voids than found in the CMASS sample.}
	\label{fig:stat_radius}
\end{figure}

\begin{figure}
	\includegraphics[scale=0.4]{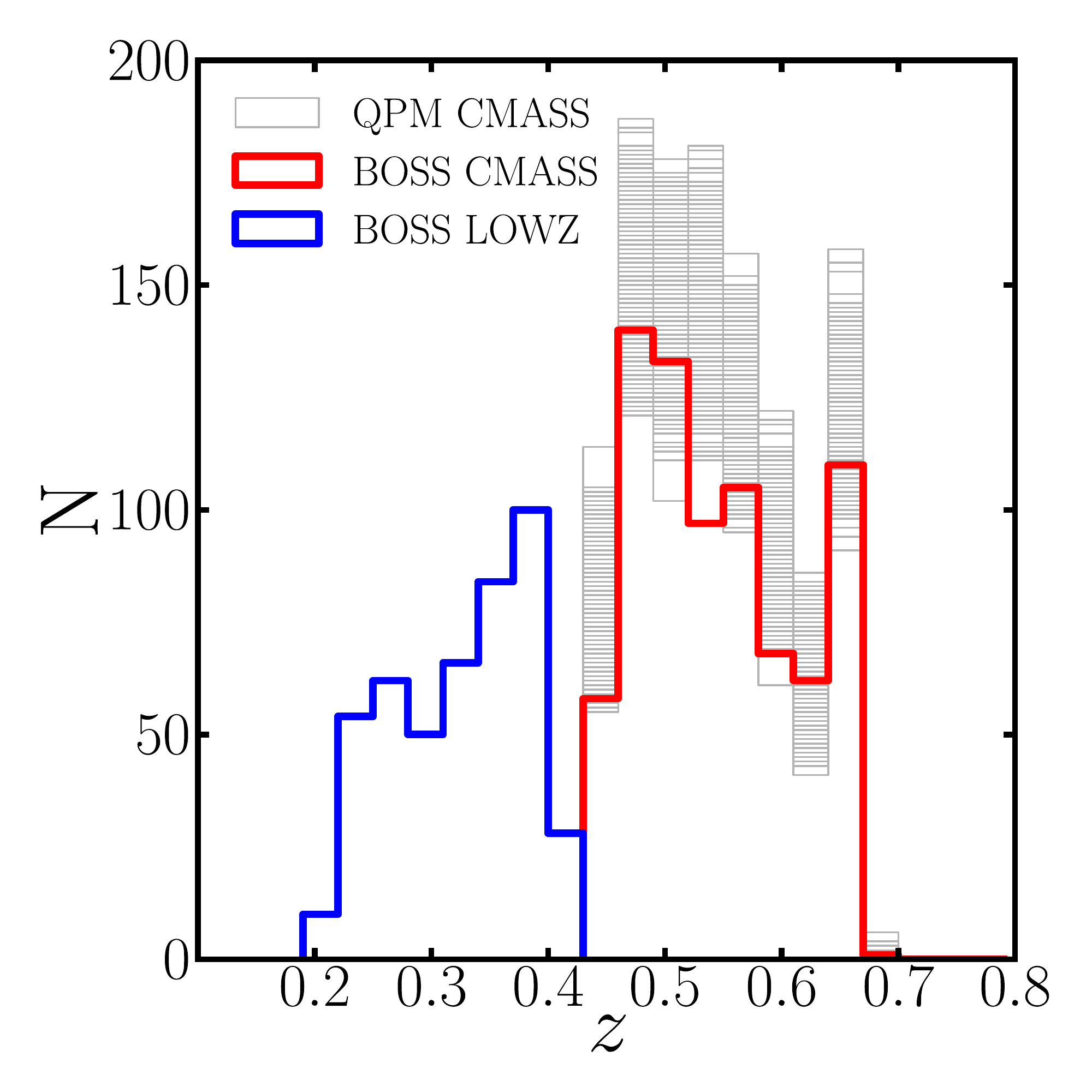}
	\caption{Distribution of void redshifts. Each gray line represents the redshift distribution of void centers from one of the 1000 QPM CMASS (North + South) mock catalogs. Results for the CMASS (North + South) and LOWZ (North + South) samples are shown by the red and blue lines, respectively.}
	\label{fig:stat_redshift}
\end{figure}

\subsection{Size and redshift distributions}

The distributions of void sizes are presented in Figure~\ref{fig:stat_radius} and the void center redshifts in Figure~\ref{fig:stat_redshift}. The measurements from the 1000 QPM CMASS mock catalogs are also plotted for comparison. In general, the measurements from the BOSS LSS catalogs agree with those from the QPM mocks. There is an overall amplitude difference between the BOSS and mock histograms such that there are 10-20\% fewer voids found in the BOSS CMASS sample than in the mean of the mocks; however, this difference is not highly significant.

The majority of the voids in this catalog have sizes ranging from $30\hmpc$ to $80\hmpc$. This is relatively large compared to the previous catalogs using the SDSS main galaxy sample \citep{Pan:2012, Sutter:2012a}, but is comparable to the previous catalogs using CMASS galaxies \citep{Sutter:2014}. The deficit of small voids is due to the sparse sampling of the galaxies in our samples. The mean galaxy separation in the CMASS sample is about $30\hmpc$, thus it is difficult to identify reliable voids smaller than that size. The void catalogs are not volume complete since the number density of galaxies depends on redshift. We naturally find relatively fewer small voids at the low and high redshift ends of each sample, where the galaxy number density is lower.

\begin{figure}
	\includegraphics[scale=0.4]{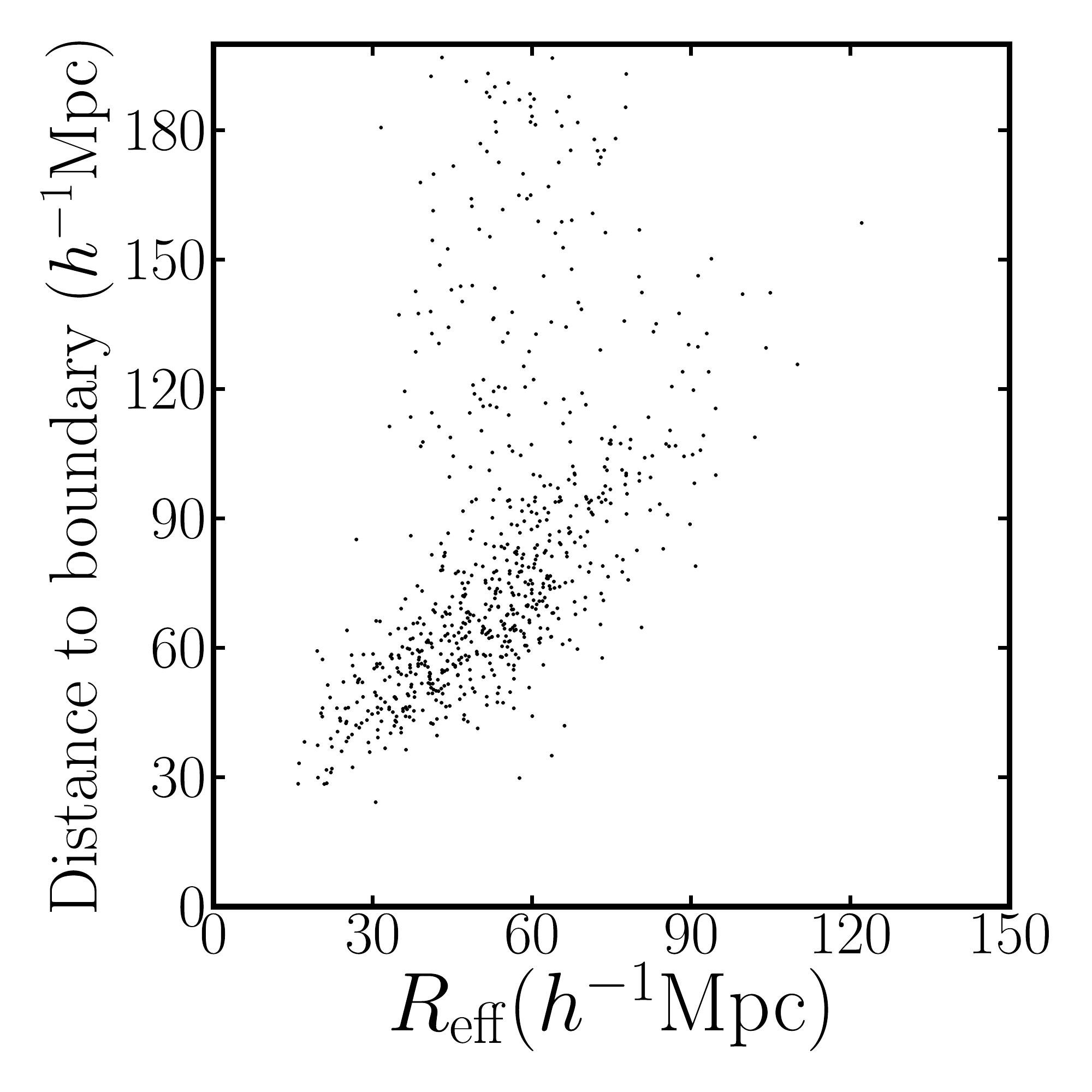}
	\caption{Distance from void center to the nearest survey boundary compared to void effective radius for the voids in the CMASS North and South samples. There is a clear correlation for voids within 100$\hmpc$ of the survey boundary, suggesting that many voids are truncated by the boundaries.}
	\label{fig:edge_vs_R}
\end{figure}

The void catalogs include the distance from each void center to its nearest survey boundary, which is calculated by finding the nearest buffer particle to the void center. Figure~\ref{fig:edge_vs_R} presents these boundary distances compared to the void sizes for the voids in the CMASS North and South samples. There is a clear correlation between the void size and the distance to the boundary for voids that are within $100\hmpc$ of the boundaries. This result suggests that many voids in this region are truncated by the survey boundaries. For science applications that require an unbiased void size distribution, it may be prudent to restrict the void samples to regions that are sufficiently far from the boundaries.

\subsection{Density profiles}

\begin{figure}
	\includegraphics[scale=0.4]{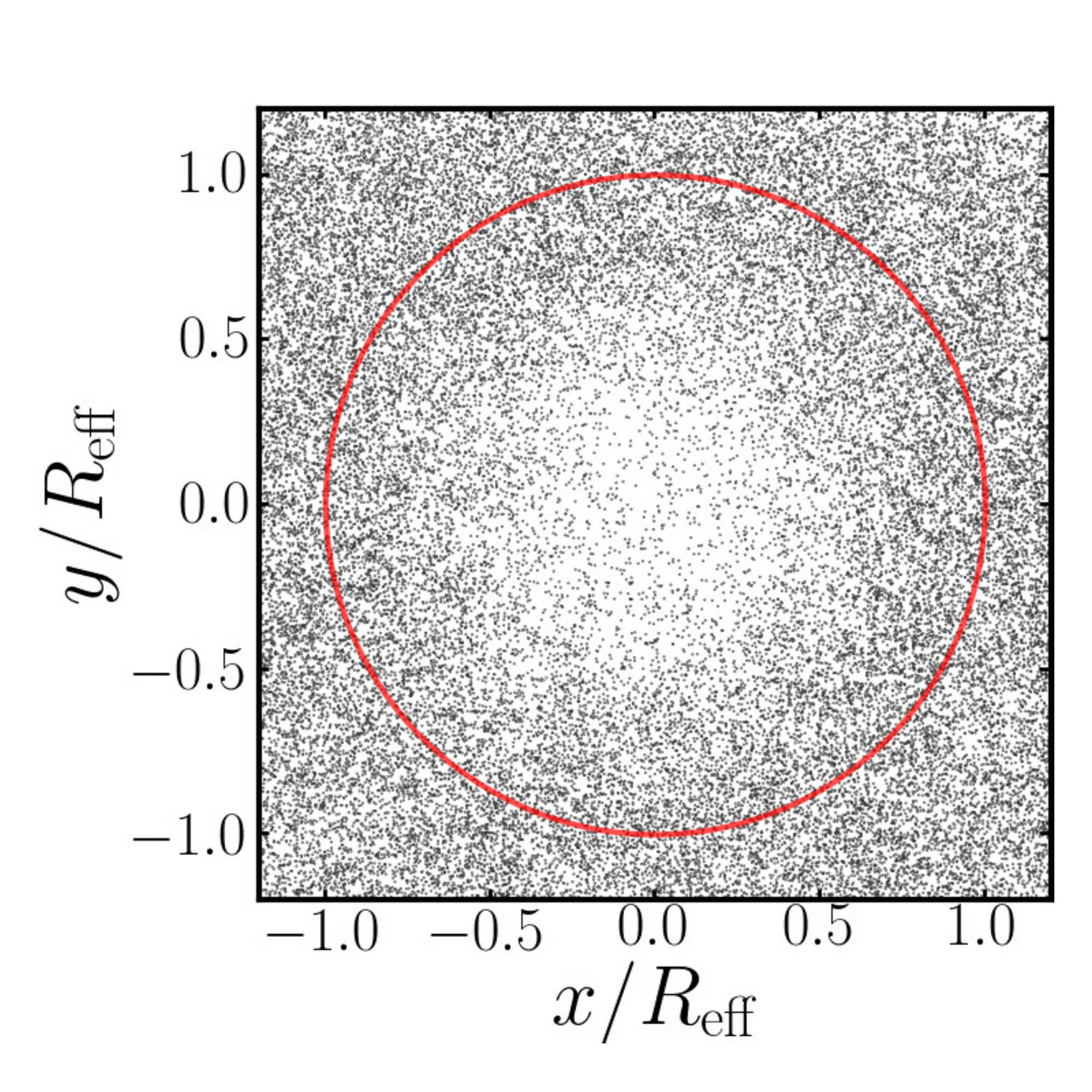}
	\caption{A slice through the stacked void using all the voids identified in the BOSS CMASS sample. Each void was rescaled by its effective radius before stacking. The slice includes all galaxies around each void center and not just the void member galaxies. The red circle shows the unit radius for reference. The central region of the stacked void is clearly underdense and roughly spherical in shape.}
	\label{fig:stacking_visual_single}
\end{figure}

Individual voids contain few galaxies and have all kinds of shapes and orientations. However, when one ``stacks" the individual voids, the composite is stable and reveals the average density structure of voids. We stack all the voids from the BOSS CMASS sample and include all the galaxies around their weighted centers (not just void member galaxies). Each void is rescaled to its effective radius before stacking. Figure~\ref{fig:stacking_visual_single} shows a slice through this stacked void. The dots show all the galaxies in a slice of the stacked void whose thickness is 0.25 times the effective radius. The stacked void looks spherical, and its central region has a low density, as expected.

\begin{figure}
	\includegraphics[scale=0.4]{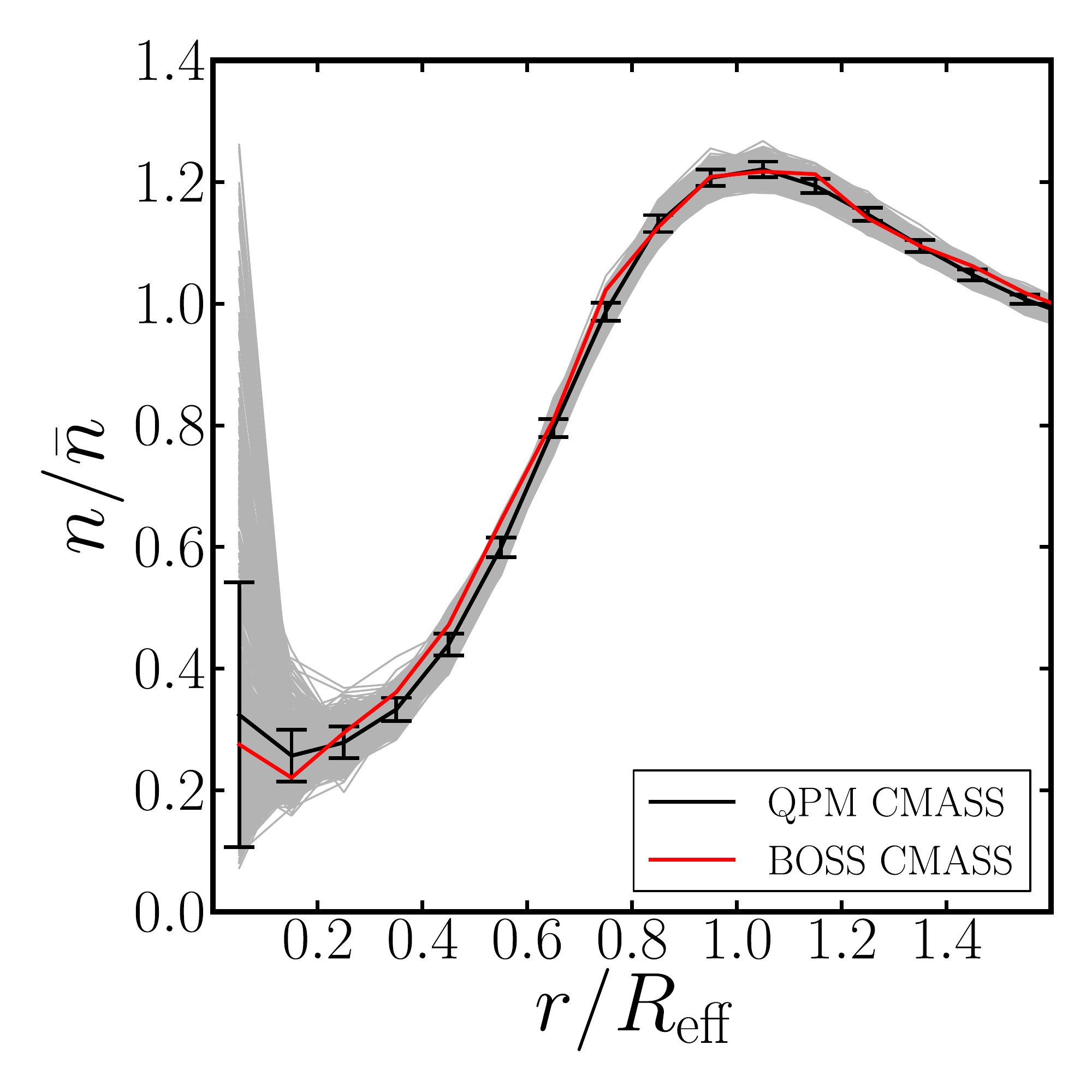}
	\caption{1-Dimensional stacked density profile from the CMASS sample. The profile is measured by calculating the density profiles for each void individually in a set of shells around each void center, scaling the densities to the mean sample density and the radii to the void effective radii, and averaging over all the voids. Each gray line represents the result for one of the 1000 QPM mock catalogs. The peak at the center is an artifact due to the way we measure the profile. The black line indicates the mean of all the mocks and error bars show the standard deviation among the mocks. The red line is the measurement from the BOSS LSS catalog. CMASS voids have central densities that are $\sim 30\%$ of the mean sample density. Moreover, the density profiles of CMASS and mock galaxies are in excellent agreement with each other.}
	\label{fig:density_profile_1d}
\end{figure}

We measure the 1-dimensional density profiles of the stacked voids by measuring the number densities $n$ in a set of shells around each void center and then scaling the number densities to the mean number density $\bar{n}$, as measured at the redshift of the void center. We then scale the radii in each void's density profile by its effective radius, and calculate the mean $n/\bar{n}$ of all the voids in our catalog. Figure~\ref{fig:density_profile_1d} presents the resulting stacked profile of the BOSS CMASS sample compared to the profiles of the QPM CMASS mocks. Since we measure the number densities for individual voids before stacking, the presence of a single galaxy in an inner shell of a small void can generate a high $n/\bar{n}$ in that shell, which produces the artificial peaks at the center. This procedure, however, ensures that the stacked density profile has the correct physical meaning. The BOSS profile agrees with the mock profiles extremely well. The density profile reveals that our ZOBOV voids have central regions with a density that is on average $\sim 30\%$ of the mean. The density peaks at a value that is 20\% higher than the mean at about one effective radius from the void center. This peak represents the walls and filaments that surround each void. The overall shape of the stacked void profile agrees with that found by previous studies, such as \citet{Sutter:2014}, \citet{Ceccarelli:2013}, and \citet{Hamaus:2014}. 

\subsection{Stellar mass distributions}

It is interesting to investigate whether galaxies living inside voids have different properties compared with galaxies living outside voids. To this end, we measure the stellar mass distributions of BOSS CMASS galaxies in different environments. The stellar masses of the galaxies are taken from the `Portsmouth SED-fit Stellar Masses' catalog, which is a value-added catalog in the SDSS data release. These stellar masses are obtained by fitting model spectral energy distributions to the observed $u$, $g$, $r$, $i$, $z$ magnitudes \citep{Fukugita:1996} of BOSS galaxies with the spectroscopic redshift determined by the BOSS pipeline, as in \citet{Maraston:2013}. There are two sets of templates available, a passive template and a star-forming template, each for both the \citet*{Salpeter:1955} and \citet*{Kroupa:2001} initial mass functions (IMF). Here we adopt the stellar masses derived from the passive template with Kroupa IMF. 

\begin{figure}
	\includegraphics[scale=0.4]{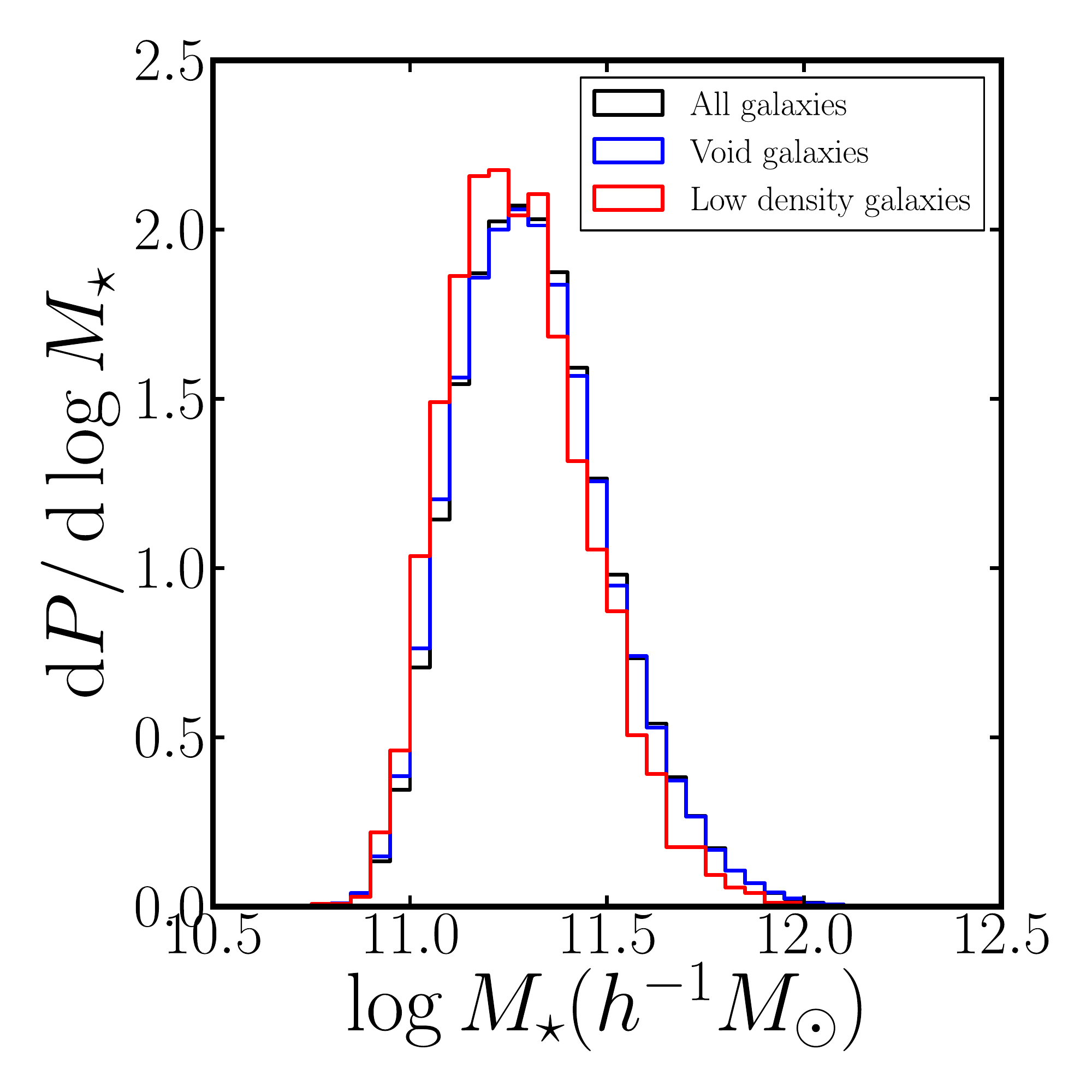}
	\caption{The stellar mass probability distribution of all BOSS CMASS galaxies (black), the void member galaxies (blue), and all galaxies with very low Voronoi density ($<0.3\bar{n}$) (red). The stellar masses of void galaxies are not appreciably different from those of all galaxies, while galaxies in low density regions are slightly less massive than all galaxies.}
	\label{fig:logmass_distr_passive_krou}
\end{figure}

Figure~\ref{fig:logmass_distr_passive_krou} presents the stellar mass distributions of all galaxies, the void galaxies, and all low-density galaxies, which we define as having Voronoi densities lower than 0.3 of the mean density. Void galaxies have stellar mass distributions that are indistinguishable from that of all galaxies. Low-density region galaxies have a distribution that is slightly shifted to lower masses, but the difference is quite small. The similarity in stellar masses is somewhat surprising as we expect low density regions to contain lower mass halos and thus less massive galaxies than high density regions. However, the BOSS CMASS sample has a fairly narrow range of stellar masses, since it only probes the high mass end of the galaxy distribution. Differences between low and high density regions can therefore not be too large. We investigate this issue further by examining the dark matter halo mass distributions of void and non-void galaxies in our QPM mocks. We find that the halo mass distributions are very similar for these different environments, which explains why we do not see a difference in the stellar mass distributions.

\section{Conclusions}
\label{s:conclusion}
We have applied the ZOBOV algorithm to the BOSS DR12 CMASS and LOWZ large-scale structure catalogs, taking into account survey boundaries, masks, and incompleteness, to construct cosmic void catalogs. These catalogs contain voids across a redshift range from $z=0.2$ to 0.7, and with effective radii spanning the range from 15 to $130 \hmpc$. The general properties of these voids, including their size and redshift distributions, as well as their stacked density profiles, are in agreement with earlier works. We have also constructed void catalogs from 1000 mock catalogs of the CMASS sample. The statistics of mock voids agree well with those of the BOSS galaxies. Finally, we have measured the stellar mass distributions of galaxies in different environments and find no significant difference between the stellar masses of void galaxies compared to all galaxies, but galaxies with very low Voronoi densities have stellar masses that are slightly lower than all galaxies. 

The cosmic void catalogs presented here are useful for many void related studies, including, but not limited to, the study of massive galaxy environments, the formation of structure on large scales, and cosmological applications such as the integrated Sachs-Wolfe effect and the Alcock-Paczy\'nski test. The void catalogs from the mock galaxy catalogs can provide information on systematic effects such as redshift distortions, and can characterize the statistical uncertainties in measured void statistics. The mock void catalogs are also useful for estimating  theoretical expectations for future surveys. Galaxy redshift surveys such as eBOSS (K. Dawson et al. 2015, in preparation), DESI \citep{Levi:2013}, Euclid \citep{Laureijs:2011} and WFIRST \citep{Spergel:2013} will produce galaxy samples in even larger volumes in the next decade, which will also greatly advance void related science. 

\acknowledgments

We thank Paul Sutter, Jennifer Piscionere, and Manodeep Sinha for valuable discussions. Q.M. and A.A.B. were supported in part by the National Science Foundation (NSF) through NSF Career Award AST-1151650. R.J.S. is supported by DOE grant DE-SC0011981. We thank the Advanced Computing Center for Research and Education (ACCRE) at Vanderbilt for providing computational resources that were used in this work. This work also used the Extreme Science and Engineering Discovery Environment (XSEDE), which is supported by National Science Foundation grant number ACI-1053575. In particular, we used the high performance computing and storage resources at the Texas Advanced Computing Center (TACC).

Funding for SDSS-III has been provided by the Alfred P. Sloan Foundation, the Participating Institutions, the National Science Foundation, and the U.S. Department of Energy Office of Science. The SDSS-III web site is http://www.sdss3.org/.

SDSS-III is managed by the Astrophysical Research Consortium for the Participating Institutions of the SDSS-III Collaboration including the University of Arizona, the Brazilian Participation Group, Brookhaven National Laboratory, Carnegie Mellon University, University of Florida, the French Participation Group, the German Participation Group, Harvard University, the Instituto de Astrofisica de Canarias, the Michigan State/Notre Dame/JINA Participation Group, Johns Hopkins University, Lawrence Berkeley National Laboratory, Max Planck Institute for Astrophysics, Max Planck Institute for Extraterrestrial Physics, New Mexico State University, New York University, Ohio State University, Pennsylvania State University, University of Portsmouth, Princeton University, the Spanish Participation Group, University of Tokyo, University of Utah, Vanderbilt University, University of Virginia, University of Washington, and Yale University.

\bibliographystyle{apj}
\bibliography{void_catalog_dr12}

\end{document}